\documentclass[10pt]{iopart}
 
\usepackage{graphicx}
\usepackage{pfnote}
\usepackage{amssymb}
\usepackage{iopams}
\usepackage{bbm}
\usepackage{bm}
\usepackage{cancel}
\usepackage{xspace} 
\usepackage{braket}
\usepackage{xcolor}   
\usepackage{calrsfs}
\usepackage{comment}
\usepackage{placeins}
\usepackage[colorlinks=true]{hyperref} 

\usepackage{units}  
 
\DeclareMathAlphabet{\pazocal}{OMS}{zplm}{m}{n} 

\newcommand{\ee}{\mathrm{e}}  
\newcommand{\ii}{\mathrm{i}} 
\newcommand*\dd{\mathop{}\!\mathrm{d}}

\renewcommand{\vec}[1]{\bm{#1}} 
\newcommand{\kel}[1]{\underline{#1}} 

\newcommand{\rre}{\mathop{\mathrm{Re}}\nolimits}
\newcommand{\iim}{\mathop{\mathrm{Im}}\nolimits}

\newcommand{\aamp}{\mathcal{A}}

\definecolor{orange}{RGB}{252,77,6}
\definecolor{brown}{RGB}{200,127,50}
\definecolor{blue}{RGB}{0,0,255}
 
\definecolor{ao(english)}{rgb}{0.0, 0.5, 0.0}

\begin{document}

\title{Phonon effects, impact ionization and power conversion in Mott photovoltaic systems}

\author{Paolo Gazzaneo$^1$ and Enrico Arrigoni$^1$}
\address{$^1$ Institute of Theoretical and Computational Physics, Graz University of Technology, 8010 Graz, Austria}
\ead{arrigoni@tugraz.at}

 
\begin{abstract}    
  
We analyze the effect of acoustic phonons on the photocurrent and the spectral characteristics of a simplified photovoltaic setup made of Mott insulating layers between two metallic leads among which a bias voltage is applied. We include acoustic phonons via the Migdal approximation and we use real-space Floquet dynamical mean-field theory to address the nonequilibrium Floquet steady-state. The so-called auxiliary master equation approach is employed as impurity solver. We find that impact ionization is only weakly affected by dissipation by acoustic phonons at low bias voltages. For higher biases instead, the \emph{Hartree shift} considerably alters the on-site energies of the Hubbard bands and suppresses the photocurrent for intermediate electron-phonon coupling strengths. Impact ionization processes play a fundamental role in enhancing the electrical output power, which decreases when electron-phonon interaction is considered.

\end{abstract}  

        
\maketitle   
 
\section{Introduction}  
\label{sec:intro}

The scientific research in the area of renewable energies and, specifically, photovoltaics is seeking for innovations able to satisfy the constantly growing energy demand. Unfortunately, the Shockley-Queisser limit~\cite{sh.qu.61} sets the maximum efficiency of conventional semiconductor photovoltaic devices and constitutes the biggest obstacle in the conversion of sunlight energy. A recent proposal to circumvent such efficiency limit is the idea of exploiting the \emph{bulk photovoltaic effect}~\cite{da.ra.22u,frid.01,ba.kr.81,pu.ro.23} and the closely related \emph{flexo-photovoltaic effect}~\cite{mi.do.18}. The latter, a strain-gradient-induced bulk photovoltaic effect, has been experimentally demonstrated in MoS$_2$~\cite{ji.ch.21} and halide perovskites~\cite{wa.sh.24} and it has been found to be orders of magnitude larger than that in non-centrosymmetric materials and in oxide perovskites. Furthermore, recent studies showed that the bulk photovoltaic effect can be substantially enhanced by electronic correlation effects in ferroelectric excitonic insulators~\cite{ka.su.21} and by collective spin dynamics in antiferromagnets~\cite{ig.wa.24}. Electronic correlation may help to overcome the Shockley-Queisser limit also in Mott insulators, thanks to the Mott gap which may be used to collect electron-hole (e-h) pairs produced by the electromagnetic radiation~\cite{mano.10,li.ch.13,gu.gu.13,co.ma.14,wa.li.15} with a possible higher efficiency due to so-called impact ionization (II)~\cite{mano.10,co.ma.14} processes.

In recent years the scientific community spent a lot of effort, from the experimental~\cite{ho.bi.16,sa.kh.23,fr.an.06,wa.mc.13,wa.bo.17} and theoretical~\cite{co.ma.14,ec.we.11,ec.we.13,we.he.14,pe.be.19,so.do.18,ka.wo.20,mano.19,ma.ev.22,wa.wa.22,ga.ma.22,ga.we.24} point of view to investigate the physics of the \emph{Mott photovoltaics} and the employment of strongly-correlated materials as efficient photovoltaic devices~\cite{wa.li.15,je.re.18}.

 At the nonequilibrium steady-state (NESS), the operating regime of a solar cell, heterostructures made of multiple correlated layers can be theoretically investigated via real-space Floquet dynamical mean-field theory (DMFT)~\cite{free.04, zl.fr.17, ma.am.15, okam.07, okam.08, ec.we.13, ec.we.14, ti.do.16, ti.so.18, qi.ho.17, qi.ho.18}. In this regard, a tentative evaluation of the efficiency, which has been estimated by Manousakis~\cite{mano.10} and evaluated by Petocchi~\cite{pe.be.19} for times up to $\sim100$ fs, would provide an important assessment for future investigations in this field. Moreover, despite the predominance of electron-electron interaction, 
electron-phonon (e-ph) scattering is an important dissipation mechanism in such systems, which competes with II as energy dissipation process of a high-energy photoexcited electron~\cite{co.ma.14}. E-ph interaction modifies the density of states of the different layers, the overlap of which plays a fundamental role
for transport properties~\cite{okam.07,okam.08}, with sizeable consequences on the photocurrent. In the present manuscript we therefore include acoustic phonons~\footnote{Acoustic phonons are more effective than optical phonons in energy dissipation. For this reason we concentrate our analysis on this case.} via the \emph{non self-consistent Migdal approximation} and analyze the photocurrent-bias voltage characteristic at the NESS for a Mott photovoltaic device. 
 
\begin{figure}
\center
\includegraphics[width=0.6\linewidth]{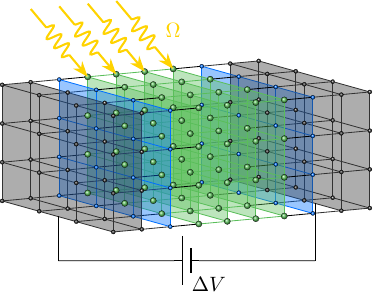}
\caption{The system considered in the manuscript. A central Mott insulating region which consists of $\mathrm{L}=4$ correlated layers, in green, is irradiated with a periodic, monochromatic, light with frequency $\Omega$. In this region acoustic phonons are also included and contribute to dissipate the harvested energy from the external driving. Two uncorrelated interfaces act as contacts, in blue, and separate the correlated region from the metallic leads, in gray. A bias voltage $\Delta V$ applied to the leads keeps them at different chemical potentials, to collect the doublon and holons created by the external driving $\Omega$.}
\label{fig:setup_voltmeter_final_fig1}
\end{figure}

The setup under consideration consists of a stack of $L=4$ Mott insulating layers~\footnote{We limit our analysis to $L=4$, since the computational effort considerably increases with larger $L$. In addition to that, we are mainly interested in the qualitative differences with respect to the single-layer case of Ref.~\cite{so.do.18,ga.ma.22}.} as in Fig.~\ref{fig:setup_voltmeter_final_fig1}, coupled to two metallic leads, acting as charge collectors to harvest the energy and kept at different chemical potentials through the application of a bias voltage. The correlated region is separated from the metallic leads by two narrowband layers representing oxide contact interfaces~\cite{as.bl.13,pe.be.19,ga.we.24}, with upper and lower Hubbard bands set as in Fig.~\ref{fig:energy_setup_final},
and driven to the NESS by an external periodic field. This driving produces e-h pairs
across the Mott gap, which are then separated by a potential gradient, which mimics the
polarization-induced electric field present in oxyde heterostructures~\cite{as.bl.13}. A Floquet steady-state is then established, with a current flowing from the left to the right lead, against the potential energy originated by the bias voltage, which contributes an electrical power gain.
Our goal is to show how dissipation by acoustic phonons changes the spectral features of the Mott insulating layers when sweeping the bias voltage and how impact ionization is thereby affected. We show how the imbalance in the number of particles between layers reflects on the photocurrent and on the  other observables when e-ph interaction is considered. We further show that II plays a substantial role for the electrical power gain and analyze how phonon dissipation limits the conversion of radiation to electrical power. In this regard, one should mention that antiferromagnetic correlations and a multiple-orbitals structure influence the efficiency of a Mott photovoltaic device as well. It has been shown in fact that the inclusion of such physics affects the dynamics and creation of photoexcited carriers~\cite{ec.we.14,pe.be.19}. These effects are, however, beyond the scope of the present analysis.
 
The rest of the manuscript is organized as follows: in Sec.~\ref{sec:Model} we describe the model under investigation, in Sec.~\ref{sec:Methods} we discuss the formalism and the mathematical tools hereby employed. We discuss the results in Sec.~\ref{sec:results} and in Sec.~\ref{sec:conclusions} we make our conclusions and final remarks.

\section{Model}
\label{sec:Model}

We consider the system shown in Fig.~\ref{fig:setup_voltmeter_final_fig1}, given by a heterostructure made up of $\mathrm{L}$ correlated layers arranged along the $z$-axis, i.e. $z\in \left\{ 1,...,\mathrm{L} \right\}$, translationally invariant in the $xy$ plane, between two metallic leads ($\rho=\mathrm{l},\mathrm{r}$). 

The Hamiltonian of the system reads
\begin{eqnarray} 
\label{eq:Hamiltonian}
\hat{H}(t) & = - \sum_{z, \langle {\mathbf r},{\mathbf r}'\rangle, \sigma}t_{z}(t) \hat{c}_{z,{\mathbf r},\sigma}^\dagger \hat{c}_{z,{\mathbf r}',\sigma}
- \sum_{\langle z, z'\rangle, {\mathbf r}, \sigma} t_{zz'} \hat{c}_{z,{\mathbf r},\sigma}^\dagger \hat{c}_{z',{\mathbf r},\sigma} \nonumber
\\ & + \sum_{z,{\mathbf r}} U_{z} \hat{n}_{z,{\mathbf r},\uparrow} \hat{n}_{z,{\mathbf r},\downarrow} + \sum_{z,{\mathbf r},\sigma} \varepsilon_z^{(0)} \hat{n}_{z,{\mathbf r},\sigma} + \hat{H}_{\mathrm{leads}} \nonumber \\ & + \hat{H}_{\mathrm{e-ph}} + \hat{H}_{\mathrm{ph}}.
\end{eqnarray} 
The operator $\hat{c}_{z,{\mathbf r},\sigma}^\dagger$ ($\hat{c}_{z,{\mathbf r},\sigma}$) is the creation(annihilation) operator of an electron on site ${\mathbf r}=(x,y)$  of  layer $z$ of spin $\sigma= \lbrace \uparrow,\downarrow \rbrace$. $\hat{n}_{z, {\mathbf r}, \sigma}=\hat{c}_{z, {\mathbf r}, \sigma}^\dagger \hat{c}_{z, {\mathbf r}, \sigma}^{\phantom\dagger}$ is the corresponding particle number operator. Here the brakets $\langle z, z'\rangle$ indicate neighboring layers along the $z$-axis while $\langle {\mathbf r}^{\phantom\dagger},{\mathbf r}'\rangle$ refers to neighboring sites belonging to the same layer. 
 
The central correlated region is under a time-periodic, homogeneous and monochromatic, electric field with frequency $\Omega$, which enters in the Hamiltonian~\ref{eq:Hamiltonian} via the Peierls substitution~\cite{peie.33} in the intralayer hopping as
\begin{equation}
\label{eq:peierls} 
t_{z}(t) = t_{z} \ e^{-\ii \frac{q}{\hbar} \left( \vec{r} - \vec{r}' \right) \cdot \vec{A}(t)},
\end{equation}
where in Eq.~\ref{eq:peierls} $\vec{A}$(t) is the time-dependent vector potential, $\hbar$ Planck's constant and $q$ the charge of the electrons. We choose $\vec{A}(t)= A(t)\vec{e}_{0}$ to point along the lattice body diagonal of a hyper-cubic lattice $\vec{e}_{0}=(1,1,\dots,1)$, where $A(t)=\frac{\hbar}{qa}\aamp\sin(\Omega t)$ and $\aamp=-\frac{qE_0a}{\hbar\Omega}$, with $E_0$ is the electric field amplitude, and $a$ the lattice spacing~\cite{ts.ok.08,mu.we.18}. The electric field is then given in the temporal gauge by $\vec{E}= -\partial_{t}\vec{A}(t) = E_0 \cos(\Omega t) \vec{e}_{0}$.
 
The second term in Eq.~\ref{eq:Hamiltonian} describes the hopping of electrons in between layers with nearest-neighbor amplitude $t_{zz'}$. The third term includes in the model the local onsite Hubbard interaction $U_z$ and the fourth one the onsite energies $\varepsilon_z^{(0)}$, the expression of which will be given in Sec.~\ref{sec:Dyson_equation}. $\hat{H}_{\mathrm{leads}}$ describes the Hamiltonian of the metallic leads, the details of which will be also given in Sec.~\ref{sec:Dyson_equation}. 

Phonons are included, in the same fashion as Ref.~\cite{ma.ga.22,ga.ma.22,ma.we.23,ma.we.24}, as acoustic branches attached to each lattice site of every layer and coupled to the electrons with the Hamiltonian 
\begin{equation}\label{eq:e-ph_ham} 
\hat{H}_{\mathrm{e-ph}} = g \sum_{z,{\mathbf r},\sigma} \hat{n}_{z,{\mathbf r},\sigma} \hat{x}_{z,{\mathbf r}},
\end{equation}
where $g$ is the e-ph coupling constant and $\hat{x}_{z,{\mathbf r}} \equiv \frac{1}{\sqrt{2}} \left( \hat{b}^{\dagger}_{z,{\mathbf r}} + \hat{b}_{z,{\mathbf r}} \right)$, with $\hat{b}^{\dagger}_{z,{\mathbf r}}$ ($\hat{b}_{z,{\mathbf r}}$) which creates (annihilates) a phonon of the acoustic branch on site ${\mathbf r}=(x,y)$ of layer $z$, whose dispersion relation is described by $\hat{H}_{\mathrm{ph}}$, discussed in Sec~\ref{sec:Dyson_equation}.

We consider a uniform Hubbard interaction $U_{z}=U$ and for the correlated region we set $t_{z}\equiv t_{\parallel}$ and $t_{zz'}= t_{\perp}$ with $t_{l,1}=t_{L,r}=v_{\rho}$ so that the hybridization between the central insulating region and the leads is symmetric on both sides of the heterostructure.  

The \emph{xy}-plane is modeled as a $d$-dimensional lattice in the limit $d \rightarrow \infty$ with the rescaling of the intralayer hopping as $t_{\parallel}=t_{\parallel}^{\ast}/(2\sqrt{d})$~\cite{tu.fr.05}. Every momentum-dependent function  depends on the electronic crystal momentum ${{\mathbf k}}$ only via $\epsilon=-2t_{\parallel} \sum_{i=1}^{d} \cos(k_i a)$ and $\overline{\epsilon}=-2t_{\parallel} \sum_{i=1}^{d} \sin(k_i a)$. Summations over the Brillouin zone are then performed using the joint
density of states (JDOS)~\cite{ts.ok.08,tu.fr.05} $\rho(\epsilon,\overline{\epsilon}) = (1/\pi t_{\parallel}^{\ast 2}) \exp[-( \epsilon^{2} + \overline{\epsilon}^{2})/t_{\parallel}^{\ast 2}]$.

In this paper we set $\hbar = k_{\mathrm{B}} = a = 1 = q$ and $t_{\parallel}^{\ast}/2=1$ as unit of energy.

\section{Methods}    
\label{sec:Methods}

\subsection{Dyson equation}\label{sec:Dyson_equation}

The lattice electron GF of central Mott insulating region obeys the Dyson equation~\cite{ti.so.18}
\begin{eqnarray}
\label{eq:FullDysonEq}
\underline{{\mathbf G}}^{-1}_{zz'}(\omega_n;\epsilon,\overline{\epsilon}) &= \underline{{\mathbf G}}^{-1}_{0,zz'}(\omega_n;\epsilon,\overline{\epsilon}) - \underline{{\mathbf \Sigma}}_{zz'}(\omega_n;\epsilon,\overline{\epsilon}) \nonumber \\ & - \underline{{\mathbf \Sigma}}_{\mathrm{e-ph},zz'}(\omega_n;\epsilon,\overline{\epsilon}).
\end{eqnarray}
The quantities in Eq.~\ref{eq:FullDysonEq} are matrices in the Floquet representation, which we use to describe the periodic NESS of the system~\cite{so.do.18,ga.ma.22,ga.we.24}. In the rest of the paper, we denote a Floquet-represented matrix as either $X_{mn}$ or $\mathbf{X}$. The underline in Eq.~\ref{eq:FullDysonEq} stands for the \emph{Keldysh} structure
\begin{equation}\label{eq:Keld-structure}
\underline{{\mathbf X}} \equiv 
\left(
\begin{array}{cc}
\mathbf{X}^{\mathrm{R}} & \mathbf{X}^{\mathrm{K}} \\
\mathbf{0}              & \mathbf{X}^{\mathrm{A}}
\end{array}
\right)
\end{equation}
with $\textbf{X}^{\mathrm{R,A,K}}$ being the \emph{retarded}, \emph{advanced} and \emph{Keldysh} components. We recall that $\textbf{X}^{\mathrm{A}}=(\textbf{X}^{\mathrm{R}})^{\dagger}$ and $\textbf{X}^{\mathrm{K}} \equiv \textbf{X}^{>} + \textbf{X}^{<}$, where $\textbf{X}^{\lessgtr}$ are the \emph{lesser} and \emph{greater} components~\cite{schw.61,keld.65,ra.sm.86,ha.ja}.

The GF which corresponds to the noninteracting terms in the Hamiltonian~\ref{eq:Hamiltonian} reads 
 \begin{eqnarray}\label{eq:non-int_InvGF} 
   & \underline{G}^{-1}_{0,mn,zz'}(\omega_n;\epsilon,\overline{\epsilon}) = \underline{g}^{-1}_{0,mn,zz'}(\omega_n;\epsilon,\overline{\epsilon}) \nonumber \\ & - \left[ v^2_{\mathrm{l}}\underline{g}_{\mathrm{l}}(\omega_n;\epsilon)\delta_{z,1} + v^2_{\mathrm{r}} \underline{g}_{\mathrm{r}}(\omega_n;\epsilon)\delta_{z,\mathrm{L}} \right] \delta_{mn}\delta_{zz'},
 \end{eqnarray}
where the shorthand notation $\omega_n \equiv \omega + n\Omega$, $n\in\mathbbm{Z}$ has been introduced. The isolated correlated layer is described by the noninteracting GF 
 \begin{eqnarray}\label{eq:inv_non-int_lat_GF_comps}
  & \left[ g^{-1}_{0}(\omega_n;\epsilon,\overline{\epsilon})) \right]^{\mathrm{R}}_{mn,zz'} = \left( \omega_n + \ii 0^{+} -\varepsilon_{z}\right)\delta_{mn}\delta_{zz'} \nonumber \\ & - t_{zz'}\delta_{mn} - \varepsilon_{mn}(\epsilon,\overline{\epsilon})\delta_{zz'},
  \end{eqnarray}
while we neglect the Keldysh component because of the presence of the leads.
 
We include the internal electric field present in oxide heterostructures, originating from the polar interfaces between the correlated layers ~\cite{as.bl.13}, via a potential drop $\Phi$ between the outermost layers of the central region, which modifies the layers' onsite energies as 
\begin{equation}\label{eq:on_site_z_dep}
\varepsilon_z =  \varepsilon_z^{(0)} + \frac{\Phi}{2} - \frac{(z-1)\Phi}{(\mathrm{L}-1)},
\end{equation}
with $\varepsilon_z^{(0)} = - U/2$ at half-filling. The off-diagonal Floquet terms in Eq.~\ref{eq:inv_non-int_lat_GF_comps} are given by the Floquet dispersion relation $\varepsilon_{mn}$ for a periodic field in a hyper-cubic lattice~\cite{ts.ok.08} 
\begin{equation}
\label{eq:Floquet_disp}
\varepsilon_{mn}(\epsilon,\overline{\epsilon}) = 
\left\{
\begin{array}{ll}
J_{m-n}(\aamp) \ \epsilon & \quad m-n:\mathrm{even} \\
\ii J_{m-n}(\aamp) \ \overline{\epsilon} & \quad m-n:\mathrm{odd},
\end{array}
\right.
\end{equation}
where $J_n$ is the $n$-th order Bessel function of the first kind with argument $\aamp$ defined as in Sec.~\ref{sec:Model}.
 
We choose metallic leads with a localized Lorentzian-shaped DOS on the surface layers (see Fig.~\ref{fig:energy_setup_final}), centered at their respective onsite energies $\varepsilon_{\rho}$, for an efficient collection of the photoexcited e-h pairs from the central region~\cite{as.bl.13,pe.be.19}. Their GF $\kel{\vec{g}}_{\rho}$ reads
\begin{eqnarray}\label{eq:bath_GFs}
g^{\mathrm{R}}_{\rho}(\omega;\epsilon) & = \left( \omega-\varepsilon_{\rho}(\epsilon) +\ii\gamma_{\rho} \right)^{-1}, \nonumber \\
g^{\mathrm{K}}_{\rho}(\omega;\epsilon) & = [g_{\rho}^{\mathrm{R}}(\omega;\epsilon) - g_{\rho}^{\mathrm{A}}(\omega;\epsilon)][1-2f(\omega,\mu_{\rho},\beta)], 
\end{eqnarray}
with $\varepsilon_{\rho}(\epsilon)=\varepsilon_{\rho} + \frac{t_{\rho}}{t^{\ast}_{\parallel}}\epsilon$ indicating the lead dispersion relation and $f(\omega,\mu_{\rho},\beta)=[\ee^{\beta(\omega-\mu_{\rho})}+1]^{-1}$ the Fermi-Dirac distribution function at inverse temperature $\beta$.

The electron self-energy (SE) $\underline{\mathbf{\Sigma}}_{zz'}$ and the 
e-ph SE $\underline{{\mathbf \Sigma}}_{\textrm{e-ph},zz'}$ are obtained from DMFT~\cite{me.vo.89,ge.ko.92,ge.ko.96} and its nonequibrium extension for driven inhomogeneous systems, i.e. real-space Floquet DMFT (F-DMFT)~\cite{ts.ok.08,po.no.99,free.04,ec.we.13,okam.07,qi.ho.17,qi.ho.18}. In this approximation, they are independent of the crystal momentum and spatially local, i.e. $\underline{{\mathbf \Sigma}}_{zz'}(\omega;\epsilon,\overline{\epsilon}) \approx \underline{\mathbf{\Sigma}}_z(\omega)\delta_{zz'}$, $\underline{{\mathbf \Sigma}}_{\textrm{e-ph},zz'}(\omega;\epsilon,\overline{\epsilon}) \approx \underline{{\mathbf \Sigma}}_{\textrm{e-ph},z}(\omega)\delta_{zz'}$. Details regarding real-space F-DMFT are given in~\ref{sec:rs_floquet_dmft}.
  
In terms of the \emph{contour times} $z,z^\prime$ and within the Migdal approximation~\cite{mu.we.15,mu.ts.17}, the e-ph SE reads~\cite{ga.ma.22,ma.we.23}: 
\begin{equation}\label{eq:backbone_e-ph_SE}
\Sigma_{\mathrm{e-ph},z}(z,z^{\prime}) = \ii g^{2} G_{\mathrm{loc},zz}(z,z^{\prime}) D_{\mathrm{ph}}(z,z^{\prime}).
\end{equation}
One can easily extract the retarded and Keldysh components of the e-ph SE from  Eq.~\ref{eq:backbone_e-ph_SE}, which are listed in Ref.~\cite{ma.ga.22}. 
  
In the current setup, away from half-filling, the \emph{Retarded} component of $\kel{\mat{\Sigma}}_{\mathrm{e-ph},z}(\omega)$ has also the contribution of the so-called \emph{Hartree shift} $\varepsilon_{\mathrm{H},z}$, which enters in the $(0,0)$-Floquet matrix element as
\begin{equation}
\label{eq:total_e_ph_SE_retarded}
\Sigma_{\mathrm{e-ph},z,00}^{\mathrm{R}}(\omega) = \varepsilon_{\mathrm{H},z} + \Sigma_{\mathrm{e-ph},z,00,\mathrm{XC}}^{\mathrm{R}}(\omega),
\end{equation} 
where $\Sigma_{\mathrm{e-ph},z,00,\mathrm{XC}}^{\mathrm{R}}(\omega)$ is the contribution coming from Eq.~\ref{eq:backbone_e-ph_SE}. The expression of $\varepsilon_{\mathrm{H},z}$ and its derivation are given in Sec.~\ref{subsec:phonons} and~\ref{sec:HS_ac_phonons} respectively. 
 
The \emph{Retarded} component of the non-interacting phonon GF $\kel{D}_{\mathrm{ph}}(t,t^{\prime})$  is given by~\cite{ao.ts.14}
\begin{equation}
\label{eq:Ph_Prop_time_ret}
D^{\mathrm{R}}_{\mathrm{ph}}(t,t^{\prime}) = -\ii \theta(t-t^{\prime}) \int \dd\omega \ \ee^{-\ii\omega\left(t-t^{\prime}\right)} A_{\mathrm{ph}}(\omega),
\end{equation} 
where $A_{\mathrm{ph}}(\omega) = (\omega/\omega^{2}_{\mathrm{ph}}) \ee^{-| \omega|/\omega_{\mathrm{ph}}}$ is the acoustic phonons' spectral function, with $\omega_{\mathrm{ph}}$ a soft cutoff frequency~\cite{pi.li.21,ga.ma.22}. The other \emph{Keldysh} components are reconstructed via the \emph{fluctuation-dissipation} theorem~\cite{st.va.13,ga.ma.22}.

\subsection{Observables} 
\label{sec:observables}  

Here we discuss the physical quantities considered for our analysis. In the present paper, we always refer to time-averaged quantities, which in the Floquet representation correspond to the $(0,0)$ matrix element, i.e.
\begin{equation}
\label{eq:av_quant_floq}
\kel{X}_{00}(\omega) = \int_{-\infty}^{+\infty} dt_{\mathrm{rel}} e^{\textrm{i} \omega t_{\mathrm{rel}}} \frac{1}{T} \int_{-\frac{T}{2}}^{\frac{T}{2}} dt_{\mathrm{av}} \kel{X}(t_{\mathrm{rel}},t_{\mathrm{av}}),
\end{equation} 
with $t_{\mathrm{rel}} = t-t^{\prime}$ and $t_{\mathrm{av}} = (t+t^{\prime})/2$ the relative and average times, respectively. The mean $j$~\footnote{As discussed in Ref.~\cite{ga.we.24}, in a steady-state situation $j_{\tilde{z},\tilde{z}+1}$ should have the same value for every $\tilde{z}$. Nevertheless, due to the limited accuracy of our AMEA impurity solver~\cite{ti.so.18}, deviations occur. For this reason, we consider the mean $j$ of the photocurrent (averaged over the bonds), together with its standard deviation $\sigma_j$.} of the steady-state photocurrent flowing from the left to the right lead~\cite{ti.so.18,okam.07,ga.we.24} and its corresponding standard deviation $\sigma_j$ are 
 \begin{eqnarray}
  j & \equiv \frac{1}{\mathrm{L}+1}\sum_{\tilde{z}=0}^{\mathrm{L}} j_{\tilde{z},\tilde{z}+1}, \nonumber \\
  \sigma_j & \equiv \sqrt{\sum_{\tilde{z}=0}^{\mathrm{L}}\frac{1}{\mathrm{L}+1}(j_{\tilde{z},\tilde{z}+1}-j)^{2}},
 \end{eqnarray} 
where $j_{\tilde{z},\tilde{z}+1}$ is the photocurrent between any of the $(\mathrm{L}+1)$ bonds of the entire Mott photovoltaic device\footnote{The bonds take into account also the one from the left lead ($\tilde{z}=0$) to the first layer ($\tilde{z}=1$) and from the last layer ($\tilde{z}=\mathrm{L}$) to the right lead ($\tilde{z}=\mathrm{L}+1$).}, given by:
\begin{equation}
\label{eq:current_junctions}
  j_{\tilde{z},\tilde{z}+1} = t_{\tilde{z},\tilde{z}+1} \int_{-\infty}^{+\infty}\frac{\dd\omega}{2\pi} \int \dd\epsilon \int \dd\overline{\epsilon} \rho\rre(J_{00,\tilde{z},\tilde{z}+1}),
\end{equation}
where $\rho$~\footnote{In Eq.~\ref{eq:current_junctions} we omit the ($\omega,\epsilon,\varepsilon$) dependence to facilitate the reading.} is the JDOS from Sec.~\ref{sec:Model} and the index $\tilde{z}$ runs over the leads' surface layers {\em and} the correlated region, i.e. $\tilde{z} \in \left\{ 0, 1, \dots,\mathrm{L} \right\}$  with $\tilde{z}=0$ indicating the left and $\tilde{z}+1=\mathrm{L}+1$ the right lead surface, the GF of which are given in Eq.~\ref{eq:bath_GFs}.
  
We characterize the spectral properties of this system via the \emph{local electron density of states} (DOS) $A_z(\omega)$ and {\em occupation function} $N_{z}(\omega)$~\footnote{The occupation function $N_z(\omega)$ tells us how many electronic states per spin are occupied with energy $\omega$ in the layer $z$ and is proportional to $G^{<}_{\mathrm{loc},00,zz}$.}, defined as
\begin{equation}\label{eq:el_spectral_function}
 A_z(\omega)=-\frac{1}{\pi}\iim[G_{\mathrm{loc},00,zz}^{\mathrm{R}}(\omega)],
\end{equation} 
and
\begin{equation}\label{eq:spec_occ}
N_z(\omega) = \frac{1}{4\pi} \left\{ \iim \left[G^{\mathrm{K}}_{\mathrm{loc},00,zz}(\omega)\right]-2\iim\left[G^{\mathrm{R}}_{\mathrm{loc},00,zz}(\omega)\right] \right\},
\end{equation}
in which $G^{\mathrm{R}}_{\mathrm{loc},00,zz}(\omega)$ and $G^{\mathrm{K}}_{\mathrm{loc},00,zz}(\omega)$ represent respectively the time-averaged \emph{retarded} and \emph{Keldysh} local GF, introduced in~\ref{sec:rs_floquet_dmft}. Another important quantity we consider is the layer-dependent double occupation 
\begin{equation} \label{eq:double_occ}
N_{\mathrm{D},z} = \langle \hat{n}_{z,\uparrow}\hat{n}_{z,\downarrow} \rangle,  
\end{equation} 
which consists of a fluctuation term 
 \begin{eqnarray}
  \Delta N_{\mathrm{D},z} 
& = \langle \left( \hat{n}_{z,\uparrow}-\langle \hat{n}_{z,\uparrow} \rangle \right) \left( \hat{n}_{z,\downarrow}-\langle \hat{n}_{z,\downarrow} \rangle \right) \rangle \nonumber \\   & 
=  N_{\mathrm{D},z}-n_{z}^{2}
 \end{eqnarray}
and a \emph{mean-field} contribution $n_{z}^{2}$, with $n_{z}=\langle \hat{n}_{z,\uparrow} \rangle = \langle \hat{n}_{z,\downarrow} \rangle$ denoting the number of particles per spin per layer~\footnote{In this work, we consider the central correlated system in a paramagnetic phase. As mentioned in Sec.~\ref{sec:intro}, antiferromagnetic correlations are important for the spreading of photoexcited carriers~\cite{ec.we.14} and should be included in the model for a possible enhancement of the efficiency of a Mott photovoltaic system.}. $n_{z}$ enters also into the \emph{total exceeding charge} in the layer $z$
\begin{equation}
\label{eq:exceeding_charge}
\Delta n_z = 2n_z -1.
\end{equation} 
A quantity related to the efficiency of the Mott photovoltaic device, which is discussed in Sec.~\ref{sec:results}, is the time-averaged absorbed power $\overline{P}_{\mathrm{abs}}$
\begin{eqnarray}
 \label{eq:avg_abs_pow}
 & \overline{P}_{\mathrm{abs}} = -E_0 \int \dd\epsilon \int \dd\overline{\epsilon} \rho(\epsilon,\overline{\epsilon}) \sum_{l=-\infty}^{\infty}\Big( J_{1-l}(\aamp) \nonumber \\ & + J_{-1-l}(\aamp)\Big) \Big[\epsilon(1-\delta_{l,0})\delta_{l,2k} + \ii\overline{\epsilon}\delta_{l,2k+1} \Big] \nonumber \\ & \sum_{z=1}^{\mathrm{L}}\int_{-\infty}^{+\infty}\frac{\dd\omega}{2\pi}G_{l,z}^{<}(\omega;\epsilon,\overline{\epsilon})
 \end{eqnarray}
in which $k\in\mathbb{Z}$. We use in~\ref{eq:avg_abs_pow} the so-called \emph{Wigner representation}, introduced in~\ref{sec:av_abs_power}, for the \emph{lesser} GF (see Sec.~\ref{sec:Dyson_equation}).

\section{Results}
\label{sec:results}     
  
\begin{figure*}
\center
\includegraphics[width=0.9\textwidth]{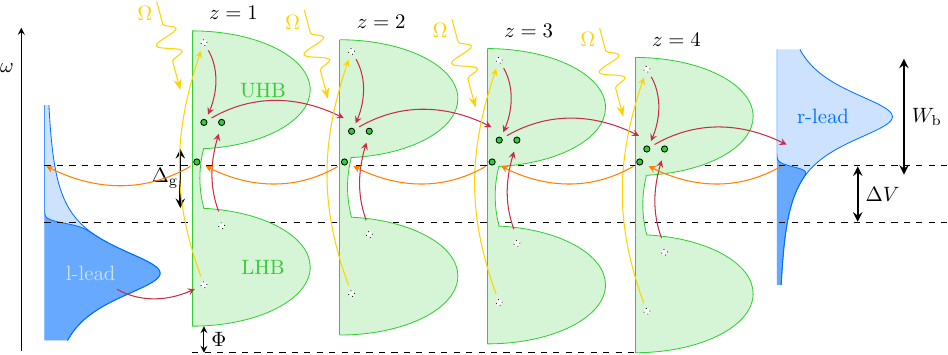}
\caption{Schematic representation of \emph{impact ionization} processes occurring in the system under analysis. Electrons are injected into the central insulating region (in green) by the left reservoir (in blue) and drained out of it by the right reservoir. $\Phi$ is the potential drop across the central region and is a linear function of the coordinate $z$ of the layers. The bias voltage $\Delta V=\mu_r-\mu_l$ applied to the metallic leads sets the occupations within them and produces a current from the right to the left of the heterostructure (orange arrows). For values of the driving frequency $\Omega \gtrsim 2\Delta_{\mathrm{g}}$, a photoexcited electron in the upper band can excite, in turn, a second electron across the gap from the lower band by transferring energy to it via Coulomb scattering. As outcome, two carriers {\em per photon} can now tunnel towards the right lead, resulting in a current from the left to the right (denoted by the red arrows), which flows against the potential energy established by the bias voltage.}
\label{fig:energy_setup_final}
\end{figure*}

We consider the setup shown  in Fig.~\ref{fig:setup_voltmeter_final_fig1}, with the local DOS landscape in Fig.~\ref{fig:energy_setup_final}, consisting of $\mathrm{L}=4$ correlated layers, a situation which has already been studied in Refs.~\cite{as.bl.13,pe.be.19,ga.we.24}. The default values for the main parameters employed in this manuscript are given in Table~\ref{tab:default_pars}~\footnote{The values of the parameters in Table~\ref{tab:default_pars} are such that $\alpha\equiv t^{\ast}_{\parallel}E_0/2\Omega^{2}<0.5$, which justifies FDSA~\cite{so.do.18,ga.ma.22,ga.we.24}.}, if not stated otherwise. 
\begin{table}[b]
  \begin{center}
\begin{tabular}{ cccccccccc }
      \hline
      \hline
        $U$ \ & $E_0$ \ & $\gamma_{\mathrm{l}/\mathrm{r}}$ \ & $v$ \ & $t_{\mathrm{l}/\mathrm{r}}$ \ & $t_{\perp}$ \ & $\Phi$ \ & $1/\beta$ \ & $\mathrm{L}$ \\
      \hline
        $12$ \ & $2$ \ & $2$ \ & $0.4$ \ & $2$ \ & $1$ \ & $2$ \ & $0.02$ \ & $4$ \\
      \hline
      \hline
    \end{tabular}
    \caption{Default values of the main parameters used in this manuscript: for simplicity we defined $v \equiv v_{\mathrm{l}/\mathrm{r}}$. As a reminder, the {\em renormalized} hopping in the correlated region is $t^{\ast}_{\parallel}=2$.}
    \label{tab:default_pars}
  \end{center}
\end{table}
We choose the leads' onsite energies $\varepsilon_{\mathrm{l}/\mathrm{r}}$ such that the left(right) lead DOS overlaps with the lower(upper) Hubbard band of the leftmost(righmost) layer with $\varepsilon_{\mathrm{l}} = \varepsilon_{1}$ and $\varepsilon_{\mathrm{r}}= -\varepsilon_{\mathrm{l}}$, see the scheme in Fig.~\ref{fig:energy_setup_final}. 
We vary the bias voltage $\Delta V=\mu_r-\mu_l$, 
so that for $\Delta V>0$ energy is harvested from the periodic driving when
an electron current flows
from the left to the right lead. We take, as convention, the current flowing from right to left as positive and the one from left to right as negative. The local DOS of the correlated layers has a gap $\Delta_{\mathrm{g}}\approx 3$, while the bandwidth of the leads $W_{\mathrm{b}} \approx 8.5$~\cite{ga.ma.22,ga.we.24}. We quantify the effect of acoustic phonons on the electronic observables via the effective e-ph coupling strength $\lambda_{\mathrm{eff}}=2g^2/W_{\mathrm{b}}\omega_{ph}$, whose derivation is found in~\ref{sec:e-ph_coup_strength}~\footnote{The values of $\lambda_{\mathrm{eff}}$ in Sec.~\ref{sec:results} are obtained taking $W_{\mathrm{b}}=8.5$, $g\in\left\{0.3,0.4,0.5\right\}$ and $\omega_{ph}\in\left\{0.05,0.1\right\}$, in line with Ref.~\cite{ga.ma.22,pi.li.21}.}.
    
In this work we are interested in assessing how II is influenced by the dissipation by acoustic phonons and how it affects the efficiency of the Mott photovoltaic system. This process, in which an electron excited to the upper Hubbard band (UHB) by a photon with energy $\Omega$ excites a second electron across the gap via electron-electron scattering, may occur if the bandwidth of the UHB, which in our setup is roughly equal to the leads' DOS, is at least twice the size of the gap $\Delta_\mathrm{g}$, i.e. $W_{\mathrm{b}} \geq 2\Delta_\mathrm{g}$~\cite{so.do.18,ga.ma.22,ga.we.24}.

 \subsection{Effects of electron-phonon interaction} 
\label{subsec:phonons}

\begin{figure*}
\center
\includegraphics[width=0.8\textwidth]{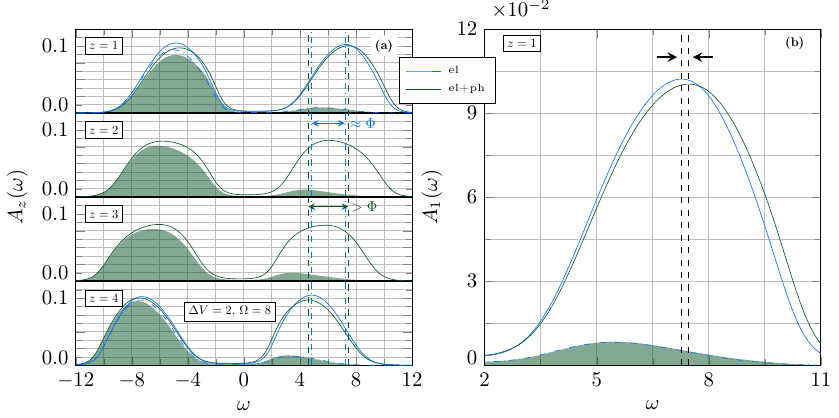}
\caption{(a) Spectral functions $A_z(\omega)$ (solid lines) and occupation functions $N_z(\omega)$ (shaded areas and dashdotted lines) for the different layers at $\Omega=8$, $\Delta V=2$ and $\lambda_{\mathrm{eff}}=0.752$. (b) Zoom of electron DOS and occupation function for the layer $z=1$. Vertical dashed and dashdotted lines in (a) highlight the separation between the center of the bands of the layers $z=1$ and $z=4$, without (blue line) and with e-ph interaction (dark green line). Vertical black lines in (b) highlight the difference in the peak positions of the bands without and with e-ph interaction. Default parameters are specified in Table~\ref{tab:default_pars}. (Here $\phi=2$, $\Delta_{\mathrm{g}} \approx 3$ and $W_{\mathrm{b}} \approx 8.5$.)}
\label{fig:Spec_func_deltaV2_Omega8_elph_fig4}
\end{figure*}

We start by addressing the effect of the e-ph interaction on the spectral and transport properties of the system.
\begin{figure*}
\center
\includegraphics[width=\textwidth]{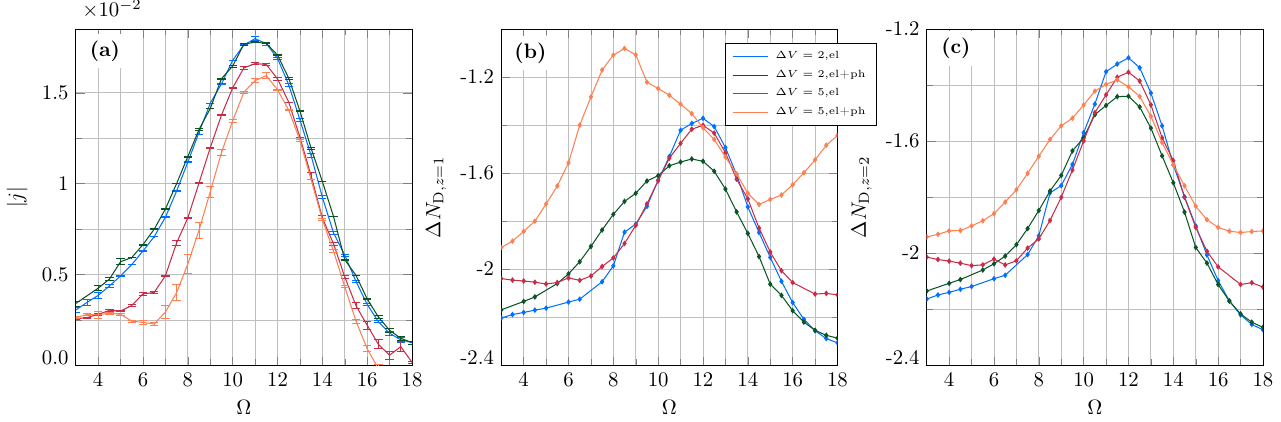}
\caption{(a) Absolute value of the time-averaged steady state current $|j|$ with error bars corresponding to $\sigma_j$ and (b)-(c) fluctuation of double occupancy $\Delta N_{\mathrm{D},z}$ for $z=1$ and $z=2$, plotted as functions of the driving frequency $\Omega$, for different values of the bias voltage $\Delta V$ without and with e-ph interaction with coupling strength $\lambda_{\mathrm{eff}}=0.752$. Default parameters are specified in Table~\ref{tab:default_pars}. Due to the PhI symmetry (see Ref.~\cite{ga.we.24}), $\Delta N_{\mathrm{D},\mathrm{L}+1-z} = \Delta N_{\mathrm{D},z}$ so the curves for $z=3$ and $z=4$ in (b)-(c) are omitted. (Here $\phi=2$, $\Delta_{\mathrm{g}} \approx 3$ and $W_{\mathrm{b}} \approx 8.5$.)}
\label{fig:j_omega_diffmuE4lorphonons_fig3} 
\end{figure*} 
Fig.~\ref{fig:j_omega_diffmuE4lorphonons_fig3} shows that the photocurrent as well as $\Delta N_{\mathrm{D},z}$ for $\Delta V=2$ in presence of the e-ph interaction (green curves) display a slight enhancement for small to intermediate $\Omega$-values. This is due to the tiny redistribution of the spectral weight from the maxima peaks towards the edges of the Hubbard bands in Figs.~\ref{fig:Spec_func_deltaV2_Omega8_elph_fig4}, which reduces slightly the gap $\Delta_{g}$, with a larger occupation of the positive-energy states next to the bottom of the UHBs for all the layers. This effect is relatively small for a single layer but becomes appreciable when considering the whole heterostructure, and leads to a small increase of the photocurrent, as already observed in Ref.~\cite{ga.ma.22}. The magnitude of the peak in $j$ at $\Delta V=2$ remains instead comparable, because of the weaker DOS suppression around the Hubbard peaks [Fig.~\ref{fig:Spec_func_deltaV2_Omega8_elph_fig4}(b)] with respect to Ref.~\cite{ga.ma.22}. 

We notice in Fig.~\ref{fig:j_omega_diffmuE4lorphonons_fig3}(a) that the kink around $\Omega\approx2\Delta_{g}$ signalizing II~\cite{ga.we.24} is not affected by the e-ph interaction. In this regard, also the fluctuation of double occupancy $\Delta N_{\mathrm{D},z}$ [green curves in Fig.~\ref{fig:j_omega_diffmuE4lorphonons_fig3}(b)-(c)] shows a change in slope for $\Omega\approx 2\Delta_{g}$, although less sharp due to the smearing by acoustic phonons.

\begin{figure*}
\center
\includegraphics[width=0.8\textwidth]{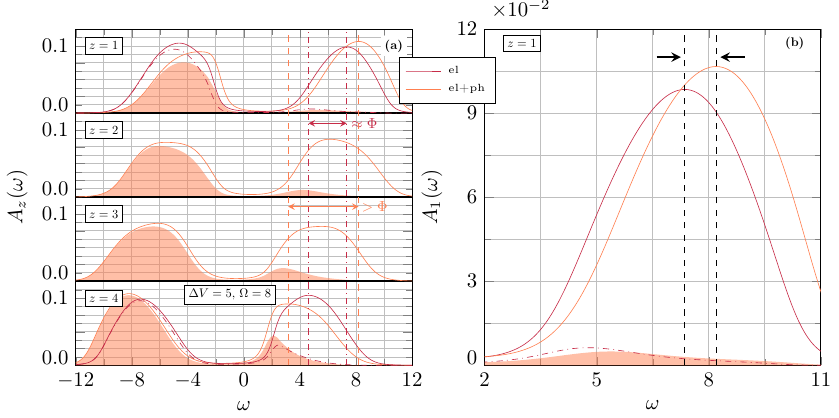}
\caption{(a) Spectral functions $A_z(\omega)$ (solid lines) and occupation functions $N_z(\omega)$ (shaded areas and dashdotted lines) for the different layers at $\Omega=8$, $\Delta V=5$ and $\lambda_{\mathrm{eff}}=0.752$. (b) Zoom of electron DOS and occupation function for the layer $z=1$. Vertical dashed and dashdotted lines in (a) highlight the separation between the center of the bands of the layers $z=1$ and $z=4$, without (red line) and with e-ph interaction (coral line). Vertical black lines in (b) highlight the difference in the peak positions of the bands without and with e-ph interaction. Default parameters are specified in Table~\ref{tab:default_pars}. (Here $\phi=2$, $\Delta_{\mathrm{g}} \approx 3$ and $W_{\mathrm{b}} \approx 8.5$.)}
\label{fig:Spec_func_deltaV5_Omega8_elph_fig5}
\end{figure*}

On the other hand, at $\Delta V=5$ the photocurrent experiences a reduction in magnitude because of the larger bias voltage which favours a current in the opposite direction~\footnote{The current at $\Omega\gtrsim17$ for $\Delta V=5$ becomes negative (not displayed in Fig.~\ref{fig:j_omega_diffmuE4lorphonons_fig3}(a)), because for such $\Omega$ the carriers' flow goes in the direction of the bias voltage.}. The presence of acoustic phonons suppresses further $j$ because the \emph{Hartree shift}, discussed in Sec.~\ref{sec:hs_effect}, moves the Hubbard bands away from each other,  and the overlap between the layers' DOS is reduced, as shown in Fig.~\ref{fig:Spec_func_deltaV5_Omega8_elph_fig5}(a).

For small $\Omega$ the available states in the layers close to the right lead are mostly filled (see $N_z(\omega)$ for $z=3,4$ in Fig.~\ref{fig:Spec_func_deltaV5_Omega8_elph_fig5}(a)). Therefore, they suppress the photocurrent and contribute to a substantial increase of $\Delta N_{\mathrm{D},z}$ (coral curve in Fig.~\ref{fig:j_omega_diffmuE4lorphonons_fig3}(b)-(c))~\footnote{The residual value of $\Delta N_{\mathrm{D},z}$ at small and large $\Omega$ (red and coral curves in Fig.~\ref{fig:j_omega_diffmuE4lorphonons_fig3}(b)-(c)) is due to the large separation between the Hubbard bands as a consequence of the Hartree shift.}. 
\subsubsection{The Hartree shift}\label{sec:hs_effect}
 
We discuss now in detail the effect of the acoustic phonons on the spectral features of the heterostructure. As described below, the most important spectral changes are due to \emph{Hartree shift} induced by the e-ph interaction.

\begin{figure*}
\center
\includegraphics[width=0.9\textwidth]{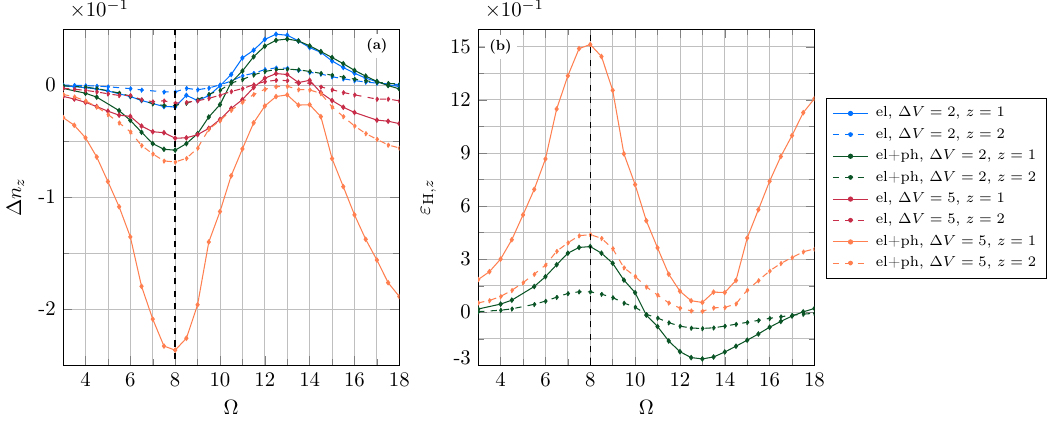}
\caption{(a) Exceeding charge $\Delta n_{z}$ without and with e-ph interaction and (b) Hartree shift $\varepsilon_{\mathrm{H},z}$ for the layers $z=1$ and $z=2$, at the e-ph coupling strength $\lambda_{\mathrm{eff}}=0.752$ for different values of the bias voltage $\Delta V$, plotted as functions of the driving frequency $\Omega$. Black dashed vertical lines in (a)-(b) highlight the values taken for the analysis of Figs.~\ref{fig:Spec_func_deltaV2_Omega8_elph_fig4}-~\ref{fig:Spec_func_deltaV5_Omega8_elph_fig5}. Due to the PhI symmetry~\cite{ga.we.24} $\Delta n_{L+1-z} = -\Delta n_{z}$ and $\varepsilon_{\mathrm{H},L+1-z}=-\varepsilon_{\mathrm{H},z}$, therefore the curves for $z=3$ and $z=4$ in (a)-(b) are omitted. Default parameters are specified in Table~\ref{tab:default_pars}. (Here $\Phi=2$, $\Delta_{\mathrm{g}} \approx 3$ and $W_{\mathrm{b}} \approx 8.5$.)}
\label{fig:Exceeding_charge_Hartree_shift_comp_phon_fig6}
\end{figure*}

We start the analysis of the total exceeding charge $\Delta n_z$ in Fig.~\ref{fig:Exceeding_charge_Hartree_shift_comp_phon_fig6}(a) first without considering e-ph interaction. For $\Omega$ up to $\sim 8$ photoexcited electrons from the first two layers $z=1,2$ tunnel to the second half of the heterostructure and leave back a charge deficiency. Such picture is confirmed by the occupation function $N_z(\omega)$ in Figs.~\ref{fig:Spec_func_deltaV2_Omega8_elph_fig4}(a)-~\ref{fig:Spec_func_deltaV5_Omega8_elph_fig5}(a). For larger $\Omega$, II is dominant and electrons in the layers $z=1,2$ build-up in doubly occupied states (see Fig.~\ref{fig:j_omega_diffmuE4lorphonons_fig3}(b)-(c)). $\Delta n_z$ is strongly influenced by the bias voltage $\Delta V$ which partially fills the UHB of the heterostructure and significantly contributes to the imbalance of $n_z$ between the layers (red and blue curves in Fig.~\ref{fig:Exceeding_charge_Hartree_shift_comp_phon_fig6}(a)), especially for those at the interfaces with respect to the ones in the bulk of the correlated region.
The number of particles per spin per layer $n_z$ is influenced also by an another factor which arises from the e-ph interaction. Away from half-filling the so-called \emph{Hartree shift}, given by the expression derived in~\ref{sec:HS_ac_phonons}
\begin{equation}
 \label{eq:hartree_shift}
 \varepsilon_{\mathrm{H},z} = -\frac{2g^2}{\omega_{ph}}\Delta n_z
\end{equation} 
modifies the on-site energies of the different layers in Eq.~\ref{eq:on_site_z_dep} and renormalizes them as function of $\Omega$. This reflects in the spectral features as follows.

At $\Delta V=2$ $\varepsilon_{\mathrm{H},z=1,2}$, is positive for $\Omega\lesssim10.5$ [Fig.~\ref{fig:Exceeding_charge_Hartree_shift_comp_phon_fig6}(b)] and the spectral functions of the layers $z=1,2$ are shifted to higher energies, as visible for $\Omega=8$ in Fig.~\ref{fig:Spec_func_deltaV2_Omega8_elph_fig4}(b), while for the layers $z=3,4$ the DOS is moved to lower energies. Overall then, the distance between the peaks of the Hubbard bands increases, as pointed out by the dark green vertical dashed lines in Fig.~\ref{fig:Spec_func_deltaV2_Omega8_elph_fig4}(a). For $\Omega\gtrsim10.5$ instead, the layers' DOS has the opposite behavior (not shown).
 
For $\Delta V=5$ $\varepsilon_{\mathrm{H},z=1,2}$ is positive in the whole $\Omega$-range considered. The single layer's DOS shift is quite large and moves the spectral functions of all layers considerably away from each other, as Fig.~\ref{fig:Spec_func_deltaV5_Omega8_elph_fig5}(a)-(b) shows~\footnote{The steep rise in $\Delta n_z$ and $\varepsilon_{\mathrm{H},z}$ for $\Delta V=5$ with e-ph interaction at $\Omega>15$ (coral curves in Fig.~\ref{fig:Exceeding_charge_Hartree_shift_comp_phon_fig6}) is due to the current flowing in the direction of the bias.}. The difference in magnitude between $\varepsilon_{\mathrm{H},z,\Delta V=2}$ and $\varepsilon_{\mathrm{H},z,\Delta V=5}$, which reflects in a much larger \emph{effective potential drop} is a direct consequence of the larger $\Delta n_z$ for higher bias voltages. $\Delta n_z$ is, in turn, strongly influenced by the spectral changes by acoustic phonons, as evident from the curves without and with e-ph interaction in Fig.~\ref{fig:Exceeding_charge_Hartree_shift_comp_phon_fig6}(a)~\footnote{In the self-consistent F-DMFT loop, changes in $\Delta n_z$ affect $\varepsilon_{\mathrm{H},z}$ and viceversa.}.

\subsection{Photocurrent-bias characteristic and device efficiency} 
\label{subsec:photcurrent_bias_characteristic}
   
In this section we analyze the power gain of the Mott photovoltaic setup. We start from the current-bias voltage chacteristic and  address the effect of phonons. 

\begin{figure*}
\center
\includegraphics[width=\textwidth]{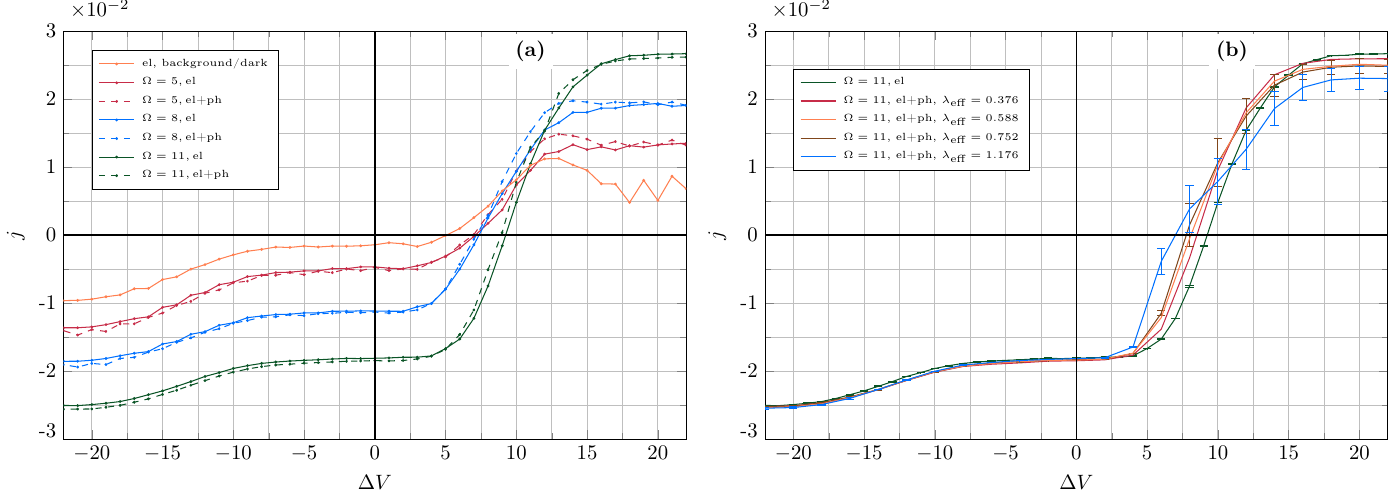}
\caption{(a) Time-averaged steady state current $j$ plotted as function of the bias voltage $\Delta V=\mu_{\mathrm{r}}-\mu_{\mathrm{l}}$ for (a) different values of the driving frequency $\Omega$ without and with e-ph interaction at $\lambda_{\mathrm{eff}}=0.212$ and (b) different e-ph coupling strengths $\lambda_{\mathrm{eff}}$ at $\Omega=11$.  The orange curve in (a) denotes the background current~\cite{so.do.18,ga.ma.22,ga.we.24} obtained without external driving and without e-ph interaction. Dark green curve in (b) without e-ph interaction is replotted from (a) for comparison. Error bars corresponding to $\sigma_j$ are not shown in (a) and only for some values of $\lambda_{\mathrm{eff}}$ in (b), for better visualization of the data.  Default parameters are specified in Table~\ref{tab:default_pars}. (Here $\phi=2$, $\Delta_{\mathrm{g}} \approx 3$ and $W_{\mathrm{b}} \approx 8.5$.)} 
\label{fig:j_delta_V_phonons_fig7} 
\end{figure*}

The photocurrent as function of the \emph{bias voltage} $\Delta V=\mu_{\mathrm{r}}-\mu_{\mathrm{l}}$ is shown in Fig.~\ref{fig:j_delta_V_phonons_fig7}. An electric power gain is obtained in the bottom right quadrant, substantially enhanced by II (for $\Omega=8,11$) with respect to the case of lower driving frequencies ($\Omega=5$). On the other hand, e-ph interaction affects only slightly the curves at small values of $\lambda_{\mathrm{eff}}$ (Fig.~\ref{fig:j_delta_V_phonons_fig7}(a)), while at higher $\lambda_{\mathrm{eff}}$ the power gain area in the II regime is clearly reduced~\footnote{Simulations at $\Omega=5,8$ for the values of $\lambda_{\mathrm{eff}}$ in Fig.~\ref{fig:j_delta_V_phonons_fig7}(b) are characterized by large values of $\Delta n_z$. This triggers an uncontrolled increase of $\varepsilon_{\mathrm{H}}$ in the self-consistent F-DMFT loop and, as consequence, simulations don't converge. For the same reason, the curves with the e-ph interaction at $\Omega=11$ show, for $\Delta V\gtrsim6$, an increase of the associated error $\sigma_j$.}, see Fig.~\ref{fig:j_delta_V_phonons_fig7}(b).

\begin{figure}
\center
\includegraphics[width=0.7\linewidth]{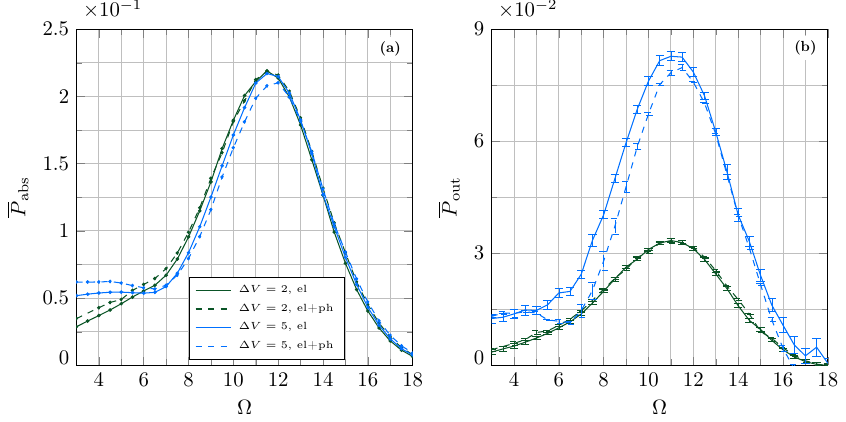}
\caption{(a) Time-averaged absorbed power $\overline{P}_{\mathrm{abs}}$ and (b) time-averaged output power $\overline{P}_{\mathrm{out}}$ with corresponding error bars, plotted as function of the driving frequency $\Omega$ for different values of the bias voltage $\Delta V=\mu_{\mathrm{r}}-\mu_{\mathrm{l}}$, without and with e-ph interaction at $\lambda_{\mathrm{eff}}=0.752$. Default parameters are specified in Table~\ref{tab:default_pars}. (Here $\Phi=2$, $\Delta_{\mathrm{g}} \approx 3$ and $W_{\mathrm{b}} \approx 8.5$.)}
\label{fig:p_abs_sweep_Omega_sweep_deltaV_fig8} 
\end{figure}

\begin{figure}
\center
\includegraphics[width=0.4\linewidth]{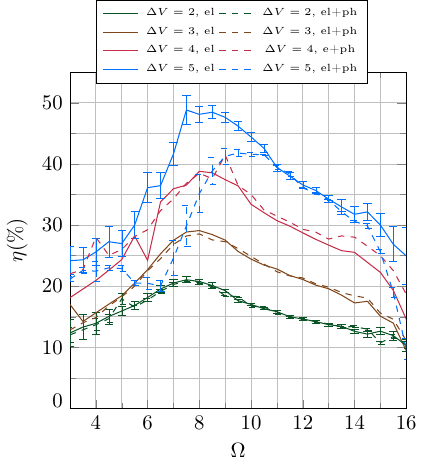}
\caption{Efficiency $\eta$ plotted as function of the driving frequency $\Omega$ for different values of the bias voltage $\Delta V=\mu_{\mathrm{r}}-\mu_{\mathrm{l}}$, without and with e-ph interaction at $\lambda_{\mathrm{eff}}=0.752$. Error bars corresponding to $\sigma_\eta$ are shown only for some values of $\Delta V$, for better visualization of the data. Default parameters are specified in Table~\ref{tab:default_pars}. (Here $\Phi=2$, $\Delta_{\mathrm{g}} \approx 3$ and $W_{\mathrm{b}} \approx 8.5$.)}
\label{fig:efficiency_fig9} 
\end{figure}

We want to study now the efficiency $\eta$ of the device, starting from the analysis of the time-averaged absorbed power $\overline{P}_{\mathrm{abs}}$ introduced in Sec.~\ref{sec:observables}. Plotted as a function of $\Omega$, this quantity shows a peak at $\Omega\approx 11$ (see Fig.~\ref{fig:p_abs_sweep_Omega_sweep_deltaV_fig8}(a)), similarly to the photocurrent $|j|$ [Fig.~\ref{fig:j_omega_diffmuE4lorphonons_fig3}(a)] and is only weakly affected by the increasing bias voltage. E-ph interaction enhances $\overline{P}_{\mathrm{abs}}$ at small $\Omega$ and suppresses it around the peak for large $\Delta V$. Due to the presence of a finite spectral weight within the gap~\footnote{A residual DOS in the gap region of the correlated layers is due to the hybridization to the leads.} a nonzero contribution of $\overline{P}_{\mathrm{abs}}$ is present also for $\Omega\lesssim\Delta_{\mathrm{g}}$. On the other hand, the time-averaged output power in Fig.~\ref{fig:p_abs_sweep_Omega_sweep_deltaV_fig8}(b) is computed as $\overline{P}_{\mathrm{out}} \equiv (|j|-|j|_{\mathrm{back}})\Delta V$, where the fictious contribution from the background current $|j|_{\mathrm{back}}$ (see orange curve in Fig.~\ref{fig:j_delta_V_phonons_fig7}(a)) has been removed.
The efficiency $\eta$, given by the ratio
\begin{equation}
 \label{eq:pow_eff}
 \eta \equiv \frac{\overline{P}_{\mathrm{out}}}{\overline{P}_{\mathrm{abs}}}
\end{equation}
provides an assessment for the power conversion quality of the photovoltaic device. $\eta$ shows a peak for $\Omega\approx8$, as one can see in Fig.~\ref{fig:efficiency_fig9}~\footnote{The uncertainties displayed in Fig.~\ref{fig:efficiency_fig9} are given by $\sigma_\eta = \frac{(\sigma_j + \sigma_{j_{\mathrm{back}}})\Delta V}{\overline{P}_{\mathrm{abs}}}$.}, confirming once more the importance of II for an efficient power conversion, in line with the results of Fig.~\ref{fig:j_delta_V_phonons_fig7}(a) (blue and dark green curves). As for the photocurrent $|j|$, $\eta$ is only weakly affected by the e-ph interaction at low $\Delta V$, in contrast to higher $\Delta V$. Even in this case, the role of II as key physical process enhancing power conversion seems to be supported by the peak at $\Omega\approx10$ displayed in Fig.~\ref{fig:efficiency_fig9} (dashed blue curve). The finite contribution of $\eta$ for $\Omega\lesssim\Delta_{\mathrm{g}}$ is due to the nonzero $\overline{P}_{\mathrm{abs}}$ and $\overline{P}_{\mathrm{out}}$ at the same frequencies, as already mentioned.
   
One should consider that the expression~\ref{eq:pow_eff} for $\eta$ is valid in the ideal case in which the electromagnetic radiation is completely absorbed by the the device region under illumination~\cite{wu.wu.16,mano.10,sp.ri.95}. In the present situation $\overline{P}_{\mathrm{abs}}$ is however some orders of magnitude smaller than $\overline{P}_{\mathrm{in}}$, the incoming power per unit surface from a perpendicular electromagnetic radiation, given by the Poynting vector $\overline{P}_{\mathrm{in}} = \frac{1}{2}c n_{\mathrm{refr}}\varepsilon_0|E_0|^2$~\footnote{Here, $c$ is the velocity of light, $n_{\mathrm{refr}}$ the refraction index of the medium and $\varepsilon_0$ the vacuum permittivity.}. This occurs because $\overline{P}_{\mathrm{in}}$ penetrates the central correlated region and is absorbed only over a large number of layers~\footnote{From Ref.~\cite{as.bl.13}, the absorption coefficient $\alpha$ in LaVO$_3$/SrTiO$_3$ gives a penetration depth of $\sim 10^{4}a$,  with $a\sim 10 \mathrm{\AA}$ the lattice spacing.}. In addition, Eq.~\ref{eq:pow_eff} holds for a monochromatic driving and therefore, in the case of a solar cell, numerator and denominator should be replaced by a weighted average over the solar spectrum.

For these reasons $\eta$ in Eq.~\ref{eq:pow_eff} should not be compared with the efficiency of a physical device, but should be rather used to qualitatively investigate promising regions of the spectrum for an efficient power conversion.
%


\section{Conclusions}
\label{sec:conclusions}  
 
In this paper we consider a simplified Mott photovoltaic setup which consists of $\mathrm{L}=4$ strongly-correlated layers under the illumination of a monochromatic, periodic electromagnetic field with driving frequency $\Omega$, connected to wideband metallic leads with narrowband layers in between. A (Floquet) steady-state photocurrent flows from the left to the right lead against the potential energy originated by the bias voltage applied to the leads' electrodes. We treat electronic correlations within real-space Floquet dynamical mean-field theory with an accurate steady-state impurity solver.

We show that dissipation by acoustic phonons affects just slightly the photocurrent when the applied bias voltage is low. This happens as consequence of the small effect that electron-phonon interaction has on the spectral features of the heterostructure layers. On the other hand, for larger values of the bias voltage, the photocurrent is suppressed by the increased distance in energy between the Hubbard bands of the different layers for intermediate electron-phonon coupling strengths. This separation is caused by the difference in the number of particles between layers which gives a large value of the \emph{Hartree shift}. 

Impact ionization processes substantially enhance the electrical power gain of the Mott photovoltaic setup with respect to the case when only direct excitations are present. Dissipation by acoustic phonons decreases such power conversion, although the reduction is not drastic.  
  
The results obtained in this paper are valid for external drivings as ultrashort pulsed lasers~\cite{mu.we.18}, with field intensities several orders of magnitude larger than those typical of the sunlight. In this work our main interest is to address the effects of electron-electron and electron-phonon interaction on transport and spectral properties of the chosen setup. For this reason, other important physical effects are missing in the present study. Future investigations which aim to quantitatively assess the efficiency of Mott photovoltaic devices should also consider those aspects. One example is the conversion of thermal energy into chemical energy, which leads to an additional chemical current~\cite{wu.wu.16}. Another one is the effect of the electron-hole pairs recombination due to the Auger effect~\cite{mano.10}, which manifests in a recombination current and reduces the output power. Also the long-range Coulomb interaction between doublons and holons has to be considered, as well as excitonic effects, which would be important to properly address the dynamics of photoexcited electron-hole pairs and might be described via self-consistent nonequilibrium GW+EDMFT schemes~\cite{go.bo.19,go.ec.19}. Moreover, charge reconstruction effects~\cite{ha.fr.12,fr.zl.07,ba.es.16} are especially important for Mott photovoltaics, due to the polar interfaces
present in the central region, the charge distribution of which needs to be properly considered. In addition to that, Refs.~\cite{zh.br.17,zh.xi.21} showed that an essential parameter for high-quality Mott photovoltaic setups is the defects concentration: one might then include impurity scattering at the level of the self-consistent Born approximation in a F-DMFT framework with low additional computational effort~\cite{ma.we.24}. Finally, magnetic effects and multi-orbital physics considerably influence the layers’ density of states~\cite{pe.be.19} and change the transport properties, as already shown in Refs.~\cite{ec.we.14,pe.be.19}.
   
\section*{Data availability statement}

The data that support the findings of this study will be openly available at the following URL/DOI: https://repository.tugraz.at/uploads/e6zzj-93j27.
   
\ack

We thank F. Petocchi and T. M. Mazzocchi for fruitful discussions. This research was funded by the Austrian Science Fund (FWF) [Grant DOI:10.55776/P33165], and by NaWi Graz. E.A. conceived the project, supervised the work and helped in several technical aspects. P.G. designed and implemented the multilayer scheme with the inclusion of acoustic phonons and produced theoretical data. The manuscript has been drafted by P.G. with contributions from E.A. The computational results presented have been obtained using the Vienna Scientific Cluster and the L-Cluster Graz.

\appendix

\section{Real-space Floquet DMFT}
\label{sec:rs_floquet_dmft}

In this Appendix we summarize the main steps of the self-consistent real-space F-DMFT~\cite{me.vo.89,ge.ko.96,ts.ok.08,ec.we.13,free.04,okam.07,ti.do.16,ti.so.18} scheme to obtain the electron and e-ph SE. In real-space F-DMFT one takes into account the electron and e-ph SE  as spatially local, i.e. $\underline{{\mathbf \Sigma}}_{zz'}(\omega;\epsilon,\overline{\epsilon}) \approx \underline{\mathbf{\Sigma}}_z(\omega)\delta_{zz'}$, $\underline{{\mathbf \Sigma}}_{\textrm{e-ph},zz'}(\omega;\epsilon,\overline{\epsilon}) \approx \underline{{\mathbf \Sigma}}_{\textrm{e-ph},z}(\omega)\delta_{zz'}$. Within this approximation, one solves for each correlated layer $z$ a (nonequilibrium) quantum impurity model with Hubbard interaction $U_z$ and onsite energy $\varepsilon_z$, with a bath hybridization function $\kel{\mat\Delta}_z(\omega)$ which has to be determined self-consistently. 
    
The steps of such real-space F-DMFT loop are~\cite{ga.we.24}:
\begin{itemize}
 \item (i) Initial guess for the electron SE $\underline{\mathbf{\Sigma}}_z(\omega)$,
 \item (ii) Computation of the local electron GF via the recursive Green's function method~\cite{ec.we.13,free.04,ti.do.16,th.ki.81,le.mu.13,ga.we.24}, i.e. $\kel{\mathbf G}_{\mathrm{loc},zz}(\omega) = \int \dd\epsilon \int \dd\overline{\epsilon} \ \rho(\epsilon,\overline{\epsilon})\underline{{\mathbf G}}_{zz}(\omega;\epsilon,\overline{\epsilon})$,
 \item (iii) Evaluation of the e-ph SE $\underline{{\mathbf \Sigma}}_{\textrm{e-ph},z}(\omega)$ via Eqs.~\ref{eq:backbone_e-ph_SE} and~\ref{eq:hartree_shift},
 \item (iv) Mapping of the problem onto a single impurity plus bath, with hybridization function $\kel{\mat{\Delta}}_z(\omega) = \kel{\mat{g}}^{-1}_{0,z,\mathrm{site}}(\omega) - \kel{\mat{G}}^{-1}_{\mathrm{loc},zz}(\omega) - \kel{\mat{\Sigma}}_z(\omega)$, in which $\kel{\mat{g}}^{-1}_{0,z,\mathrm{site}}(\omega)$ is defined as in Eq.~\ref{eq:inv_non-int_lat_GF_comps} with $\varepsilon_{mn}(\epsilon,\overline{\epsilon})=0$ and $t_{zz'}=0$,
 \item (v) Solution of the nonequilibrium many-body impurity problem, which provides the new $\kel{\mat{\Sigma}}_z(\omega)$,
\item (vi) Insertion of electron and e-ph SEs into step (ii) and iteration of steps (ii)-(vi) until convergence.
\end{itemize}
In step (iv) we consider only the $(0,0)$-Floquet matrix element of all the quantities in the expression for $\kel{\mat{\Delta}}_z(\omega)$, because we adopt the Floquet-diagonal self-energy approximation (FDSA)~\cite{so.do.18,ga.ma.22,ga.we.24} for $\underline{\mathbf{\Sigma}}_z(\omega)$ and $\underline{{\mathbf \Sigma}}_{\textrm{e-ph},z}(\omega)$, whereby the nondiagonal Floquet indices of such quantities are neglected. The other diagonal components of the SE are obtained by making use of the property $\kel{\Sigma}_{mm}(\omega) = \kel{\Sigma}_{00}(\omega+m\Omega)$. Thanks to the PhI symmetry of the system~\cite{ga.we.24}, we solve the many-body problem in step (v) only for half of the correlated layers, via the auxiliary master equation approach (AMEA)~\cite{ar.kn.13,do.nu.14,do.ga.15,do.so.17,ar.do.18,ma.we.23,we.lo.23,we.zi.24}.

\section{The \emph{Hartree shift} for acoustic phonons}
\label{sec:HS_ac_phonons}

The starting point for the derivation of the \emph{Hartree shift} $\varepsilon_{\mathrm{H},z}$ per spin per layer is the expression for the singular part of $\Sigma_{\mathrm{e-ph},z}^{\mathrm{R}}(z)$, in which $z$ as subscript indicates the layer index and $z$ in the round brackets the Keldysh contour time argument of the SE. This quantity reads
\begin{equation}
\label{eq:sing_part_e_ph_SE_contour}
\Sigma_{\mathrm{e-ph},z,\mathrm{H}}(z) = - 2\ii g^{2} \int_{\gamma} dz' D_{\mathrm{ph}}(z,z')G_{\mathrm{loc},zz}(z';{z'}^{+})
\end{equation} 
and its \emph{Retarded} component is
\begin{eqnarray}
\Sigma_{\mathrm{e-ph},z,\mathrm{H}}^{\textrm{R}}(t) & = - 2\ii g^{2} \int_{t_0}^{\infty} dt' D_{\mathrm{ph}}^{\textrm{R}}(t,t')G_{\mathrm{loc},zz}^{<}(t';{t'}) \nonumber \\ & = 2 g^{2}n_z \int_{t_0}^{\infty} dt' D_{\mathrm{ph}}^{\textrm{R}}(t,t'), \label{eq:sing_part_e_ph_SE_retarded}
\end{eqnarray}
where in the last line we use that $G_{\mathrm{loc},zz}^{<}(t';{t'}) = \ii n_z$ at the NESS. The starting point $t_0$ of the Keldysh contour may be taken as $t_0=0$ or $t_0=-\infty$ with no influence on the final result. Using Eq.~\ref{eq:Ph_Prop_time_ret} for $D_{\mathrm{ph}}^{\textrm{R}}(t,t')$ and $t_0=0$, after some standard integrations we arrive to
\begin{equation}
\label{eq:HS_ac_time}
\Sigma_{\mathrm{e-ph},z,\mathrm{H}}^{\textrm{R}}(t) = -\frac{4g^2 n_z}{\omega_{ph}^3} \frac{t^2\omega_{ph}^4}{1+t^2\omega_{ph}^2}.
\end{equation} 
Taking the limit for $t\rightarrow\infty$, one obtains the NESS expression for the \emph{Hartree shift}:
\begin{equation}
\label{eq:HS_ac_final}
 \varepsilon_{\mathrm{H},z} = \lim_{t\rightarrow\infty}\Sigma_{\mathrm{e-ph},z,\mathrm{H}}^{\textrm{R}}(t) = -\frac{2g^2}{\omega_{ph}}\Delta n_z,
\end{equation} 
in which we use Eq.~\ref{eq:exceeding_charge}. This quantity, which vanishes at half-filling, has to be computed self-consistently in the DMFT loop~\footnote{The exceeding charge depends on the number of particles which is, in turn, computed self-consistently at every DMFT iteration.}.

\section{The effective electron-phonon coupling strength}
\label{sec:e-ph_coup_strength}

The effective e-ph coupling strength used in this paper is derived from the \emph{so-called} \emph{Eliashberg function} \cite{gius.17} in the \emph{quasielastic approximation}:
\begin{equation}
 \label{eq:eliash_simpl}
 \alpha^{2}F_{\mathbf k}(\varepsilon,\omega)= \sum_{\mathbf q} \delta(\omega-\omega_{\mathbf q}) \sum_{\mathbf k'}|g_{\mathbf k',\mathbf k}^{\mathbf q}|^{2}\delta(\varepsilon-\varepsilon_{\mathbf k'}),
\end{equation} 
in which $\mathbf q$ and $\mathbf k, \mathbf k'$ are the phonon and electron crystal momenta and $g_{\mathbf k',\mathbf k}^{\mathbf q}$ is the general e-ph matrix element. This function provides the information relative to the e-ph coupling of the electronic state $\mathbf k$. Given the e-ph interaction Hamiltonian in Eq.~\ref{eq:e-ph_ham}, this expression simplifies to
\begin{equation}
 \label{eq:eliash_simpl_2}
 \alpha^{2}F(\varepsilon,\omega)= g^{2} \sum_{\mathbf q} \delta(\omega-\omega_{\mathbf q}) \sum_{\mathbf k'} \delta(\varepsilon-\varepsilon_{\mathbf k'}).
\end{equation}
We have now the product of the \emph{phonon} and \emph{electron} spectral functions separately. For the former we use the acoustic phonons' DOS $A_{\mathrm{ph}}(\omega) = (\omega/\omega^{2}_{\mathrm{ph}}) \ee^{-| \omega|/\omega_{\mathrm{ph}}}$ from Sec.~\ref{sec:Dyson_equation}, which leads to
\begin{equation}
 \label{eq:eliash_simpl_3}
 \alpha^{2}F(\varepsilon,\omega)= g^{2} A_{\mathrm{ph}}(\omega)A(\varepsilon)\approx g^{2} \frac{A_{\mathrm{ph}}(\omega)}{W_{\mathrm{b}}},
\end{equation}
where we approximate the electron DOS $A(\varepsilon)$ with the inverse of the Hubbard bands' bandwidth $W_{\mathrm{b}}$, introduced in Sec.~\ref{sec:results}. The strength of e-ph coupling is then encoded in the dimensionless parameter~\cite{mu.we.16_1} 
\begin{equation}
 \label{eq:e-ph_coup_par}
 \lambda_{\mathrm{eff}} = 2 \int_{0}^{\infty} \frac{d\omega}{\omega} \alpha^{2}F(\omega) = \frac{2g^{2}}{W_{\mathrm{b}}\omega_{\mathrm{ph}}},
\end{equation} 
in which $\alpha^{2}F(\omega)$ from Eq.~\ref{eq:eliash_simpl_3} depends now only on $\omega$. Finally, we notice that the \emph{Hartree shift} and $\lambda$ are related by:
\begin{equation}
 \label{hartree_shift_lambda}
 \varepsilon_{\mathrm{H},z} = -\lambda_{\mathrm{eff}} W_{\mathrm{b}} \Delta n_z.
\end{equation}  

\section{Derivation of the time-averaged absorbed power}
\label{sec:av_abs_power}

In this Appendix we derive the time-averaged absorbed power $\overline{P}_{\mathrm{abs}}$ introduced in Sec.~\ref{sec:observables} of the main text.

The time-dependent absorbed power $P_{\mathrm{abs}}(t)$ originates from the oscillating current $\vec{j_{\parallel}}(t)$ along the lattice body diagonal of the $d$-dimensional correlated layers, in the direction of the external electric field $\vec{E}(t)$. 

We start from the expression of $P_{\mathrm{abs}}(t)$ as the classical expectation value
\begin{equation}
 \label{eq:abs_pow_time}
 P_{\mathrm{abs}}(t) = \frac{\dd E_{\mathrm{abs}}}{\dd t} = \Big\langle \frac{\dd\hat{H}_{\mathrm{H}}(t)}{\dd t} \Big\rangle.
\end{equation}
The time-dependence in $\hat{H}_{\mathrm{H}}(t)$~\footnote{The subscript $\mathrm{H}$ stays for the \emph{Heisenberg representation}.} appears only via the vector potential $\vec{A}(t)$ in the intralayer hopping term $t_z(t)$~\ref{eq:peierls}, introduced in Sec.~\ref{sec:Model}. For this reason, one obtains the general result~\cite{ao.ts.14}
\begin{equation}
 \label{eq:abs_pow_time_chain_rule}
 P_{\mathrm{abs}}(t) = \Big\langle \frac{\partial\hat{H}_{\mathrm{H}}(t)}{\partial\vec{A}(t)}\cdot \frac{\partial\vec{A}(t)}{\partial t} \Big\rangle = \vec{j_{\parallel}}(t)\cdot\vec{E}(t),
 \end{equation}
in which the current $\vec{j_{\parallel}}(t)$ is given by~\cite{ao.ts.14,mu.we.18}
\begin{equation}
 \label{eq:current_parallel}
 \vec{j}_{\parallel}(t) = -  \Big\langle \frac{\partial\hat{H}_{\mathrm{H}}(t)}{\partial\vec{A}(t)} \Big\rangle = \frac{2\ii}{N}\sum_{z=1}^{\mathrm{L}}\sum_{\vec k}\vec{v}_{\vec{k}}(t) G_{z,\vec{k}}^{<}(t,t),
\end{equation}
where $N$ is the number of lattice sites of the single $d$-dimensional correlated layer and $n_{z,\vec{k}}(t) = -\ii G_{z,\vec{k}}^{<}(t,t)$.

In the NESS under study, we consider the time-averaged absorbed power, the expression of which is
\begin{eqnarray}
 & \overline{P}_{\mathrm{abs}} = \frac{1}{T}\int_{-T/2}^{T/2}\dd t P_{\mathrm{abs}}(t) \nonumber \\ & = \frac{2\ii}{N}\sum_{z=1}^{\mathrm{L}}\sum_{\vec k}\frac{1}{T}\int_{-T/2}^{T/2}\dd t \vec{v}_{\vec{k}}(t)\cdot\vec{E}(t) \nonumber \\ & \sum_{l=-\infty}^{\infty}e^{-\ii l\Omega t}\int_{-\infty}^{+\infty}\frac{\dd\omega}{2\pi}G_{l,z,\vec k}^{<}(\omega),
 \label{eq:avg_abs_pow_interm}
\end{eqnarray}
where we use for the \emph{lesser GF} the Wigner representation~\cite{ts.ok.08}
\begin{equation}
 \label{eq:wign_rep}
    \kel{G}_{l}(\omega)= \int_{-\infty}^{+\infty} dt_{\mathrm{rel}} e^{\textrm{i} \omega t_{\mathrm{rel}}} \frac{1}{T} \int_{-\frac{T}{2}}^{\frac{T}{2}} dt_{\mathrm{av}} e^{\textrm{i} l \Omega t_{\mathrm{av}}} \kel{G}(t_{\mathrm{rel}},t_{\mathrm{av}}),
\end{equation}
with $t_{\mathrm{rel}} = t-t^{\prime}$ and $t_{\mathrm{av}} = (t+t^{\prime})/2$ defined in Sec.~\ref{sec:observables}. The scalar product in~\ref{eq:avg_abs_pow_interm} reads
\begin{eqnarray} 
 \label{eq:vdotE}
 \vec{v}_{\vec{k}}(t)\cdot\vec{E}(t) & = E_0\cos(\Omega t)\Big\{-\overline{\epsilon}\cos\Big[\aamp\sin(\Omega t)\Big] \nonumber \\ & +\epsilon\sin\Big[\aamp\sin(\Omega t)\Big]\Big\}
\end{eqnarray}
with the quantities $\aamp$,$\epsilon$,$\overline{\epsilon}$ defined in Sec.~\ref{sec:Model}. After some algebra, one finally obtains for the Wigner and layer resolved integrand in~\ref{eq:avg_abs_pow_interm}
\begin{eqnarray} 
 \label{eq:avg_abs_pow_zl}
 \overline{P}_{\mathrm{abs},l,z}(\omega) & = -E_0 \int \dd\epsilon \int \dd\overline{\epsilon} \rho(\epsilon,\overline{\epsilon}) \Big( J_{1-l}(\aamp) + J_{-1-l}(\aamp)\Big) \nonumber \\ & \Big[\epsilon(1-\delta_{l,0})\delta_{l,2k} + \ii\overline{\epsilon}\delta_{l,2k+1} \Big] G_{l,z}^{<}(\omega;\epsilon,\overline{\epsilon}),
\end{eqnarray}
in which $k\in\mathbb{Z}$, $J_n$ is the $n$-th order Bessel function of the first kind and the integration over the JDOS $\rho(\epsilon,\overline{\epsilon})$ is used (see Sec.~\ref{sec:Model}), which leads to Eq.~\ref{eq:avg_abs_pow} in Sec.~\ref{sec:observables}. 
 
\newpage

\bibliographystyle{iopart-num}  
\bibliography{reference_database_copied}

\end{document}